\documentclass[12 pt]{article}

\usepackage{graphics}
\usepackage{graphicx}
\usepackage{setspace}

\newcommand{\NC}{\newcommand} % new command

\newcommand{\RC}{\renewcommand}

\RC{\L}{\left}  % left delimiter
\NC{\R}{\right} % right delimiter
\NC{\I}{\item} % item

\NC{\FN}{\footnote}  % footnote
\NC{\HS}{\hspace}    % horizontal space
\NC{\UL}{\underline} % underline
\NC{\VS}{\vspace}    % vertical space

\NC{\C}{\cite}  % citation
\NC{\N}{\label} % name

\NC{\iv}[1]{\L( #1 \R)} % independent variables

\NC{\ML}{\times} % multiply 
\NC{\IP}{\cdot}  % inner (dot) product
\NC{\OP}{\times} % outer (cross) product

\NC{\V}[1]{\UL{#1}} % vector
\NC{\D}[1]{\UL{\UL{#1}}} % dyadic

\NC{\im}{\textrm{Im}} % imaginary part

\NC{\PV}{\V{r}} % position vector

\NC{\EF}{\V{E}} % macroscopic electric field
\NC{\BF}{\V{B}} % macroscopic magnetic field
\NC{\DF}{\V{D}} % macroscopic electric induction field
\NC{\HF}{\V{H}} % macroscopic magnetic induction field

\RC{\AE}{\UL{{\cal E}}}    % electric field amplitude
\NC{\AF}{\omega}   % angular frequency
\NC{\WN}{k}        % wavenumber
\NC{\WV}{\T{k}{1}} % wavevector
\NC{\PH}{\phi}     % carrier phase

\NC{\ct}{\ldots} % to be continued (ellipsis)
\NC{\tr}[1]{\mathrm{#1}} % roman text in mathematical expressions

\NC{\oo}{\infty} % infinity
\NC{\ip}{\cdot} % inner product
\NC{\op}{\times} % outer product
\NC{\nd}{\frac} % fractions
\NC{\ml}{\times} % multiply
\NC{\mv}{\textrm{Max}} % maximum value
\NC{\re}{\textrm{Re}} % real part
\NC{\sr}{\sqrt} % square root
\NC{\su}{\vert} % substitute
\NC{\gt}{\rightarrow} % goes to in limit
\NC{\cv}{\ast} % convolution operator

\NC{\av}[1]{\L| #1 \R|} % absolute value
\NC{\pd}[2]{\partial_{#2} #1} % partial derivative
\NC{\ci}[2]{\L[ #1, #2 \R]} % closed interval
\NC{\oi}[2]{\L( #1, #2 \R)} % open interval

\NC{\grad}{\nabla} % gradient operator
\NC{\divg}{\nabla \ip} % divergence operator
\NC{\curl}{\nabla \op} % curl operator

\NC{\DD}[1]{\delta \iv{#1}} % Dirac delta function
\NC{\US}[1]{{\cal U} \iv{#1}} % unit step function

\NC{\mr}[1]{\L[ #1 \R]} % matrix representation of vectors and dyadics

\NC{\UV}[1]{\V{u}_{#1}} % unit vector
\NC{\NV}{\V{0}} % null vector
\NC{\ND}{\D{0}} % null dyadic
\NC{\ID}{\D{I}} % identity dyadic
\NC{\NM}{\mr{\D{0}}} % null matrix
\NC{\IM}{\mr{\D{I}}} % identity matrix

\NC{\FQ}{\omega} % angular frequency

\NC{\PT}{\iv{\PV, t}} % dependence on position vector and time
\NC{\PW}{\iv{\PV, \FQ}} % dependence on position vector and angular frequency
\NC{\XYZW}{\iv{x, y, z, \FQ}} % dependence on z position and angular frequency
\NC{\ZB}[1]{z_{#1}} % STF z-axis boundary
\NC{\YB}[1]{y_{#1}} % STF y-axis boundary

\NC{\Co}{c_{o}} % speed of light in vacuum
\NC{\Eo}{\epsilon_{o}} % permittivity of vacuum
\NC{\Mo}{\mu_{o}} % permeability of vacuum
\NC{\No}{\eta_{o}} % impedance of vacuum

\NC{\PF}{\V{P}} % polarization field
\NC{\SF}{\V{S}} % Poynting vector field
\NC{\Sz}{S_{z}} % axial component of Poynting vector

\NC{\EE}{\D{\epsilon}_{r}} % relative permittivity dyadic
\NC{\ER}{\D{\epsilon}_{ref}} % reference permittivity dyadic
\NC{\SD}[1]{\D{\chi}_{#1}} % susceptibility dyadic
\NC{\RD}[1]{\D{S}_{#1}} % rotation dyadic
\NC{\SDr}{\D{\chi}_{ref}} % reference susceptibility dyadic

\NC{\AR}{\alpha} % angle of rise
\NC{\SH}{h} % structural handedness parameter

\NC{\LT}[1]{\chi_{#1}} % Lorentzian properties function
\NC{\OS}[1]{p_{#1}} % oscillator strength
\NC{\NP}{p_{nl}} % nonlinear parameter
\NC{\AP}[1]{N_{#1}} % absorption parameter
\NC{\RW}[1]{\lambda_{#1}} % resonance wavelength
\NC{\RF}[1]{\omega_{#1}} % resonance angular frequency

\begin{document}

\begin{center}
{\large Thin-Film Metamaterials called Sculptured Thin Films} \\
\vskip 8 pt
 {Akhlesh Lakhtakia$^1$  and  Joseph B. Geddes III$^2$ }\\
  
  \vskip 6pt
     
  $^1$Department of Engineering Science and Mechanics, Pennsylvania
  State University, University Park, PA 16802, USA\\
  
  \vskip 6pt
   $^2$Beckman Institute, University of Illinois at Urbana--Champaign,
  Urbana, IL 61801, USA\\

\end{center}

\begin{abstract}
Morphology and performance are conjointed attributes of metamaterials, of which
sculptured thin films (STFs) are examples.
STFs are assemblies of nanowires that can be fabricated from many different materials, typicially via physical vapor deposition onto rotating substrates. The curvilinear--nanowire morphology of STFs is determined by the substrate motions during fabrication. The optical properties, especially, can be tailored by varying the morphology of STFs. In many cases prototype devices have been fabricated for various optical, thermal, chemical, and biological applications. 
\end{abstract}

% Throughout we stress the connections that STFs engender between phase, length, and time at the nanoscale for optics, as well applications of the designed breaking of symmetries in STFs. 

%------------------------------%
% Body:
%------------------------------%

% panoramic view with select examples

% theme: phase, length, time
% theme: symmetry breaking

% time domain
% slanted chiral
% bio uses

\section{Introduction} \label{S: Introduction}

The emergence of metamaterials at the end of the 20$^{th}$ century heralded a major motivational shift in research on materials and coincided with the ramping up of nanotechnology \cite{A.Lakhtakia-2007(P)}. Materials researchers began to consider the design of 
composite materials, called metamaterials, to perform more than role each in specific environments.  
Among optics researchers today, the term \emph{metamaterial} is often taken to mean  {a material with negative refractive index}
\cite{T.G.Mackay-2009(P)}, but there is much more to metamaterials than that. In the year 2000, Rodger 
Walser \cite{R.Walser-2003(C)}
coined this term for certain types of artificial materials, and later formally defined metamaterials as ``macroscopic composites having a manmade, three-dimensional, periodic cellular architecture designed to produce an optimized combination, not available in nature, of two or more responses [emphasis in the original] to specific excitation.Ó We can relax the requirements of periodicity today, though not of cellularity. Indeed, cellularity in morphology engenders multifunctional performance.

Sculptured thin films (STFs) \cite{A.Lakhtakia-2005(B)} exemplify metamaterials. An STF is an assembly of nanowires typically grown by physical vapor deposition, whose bent and twisted forms are engineeered via the growth process. As a result of the flexibility in controlling the evolving nanostructure of the films during fabrication, their performance characteristics can be engineered.

\begin{figure}[t]
  \begin{center}
    \scalebox{0.4}{\includegraphics{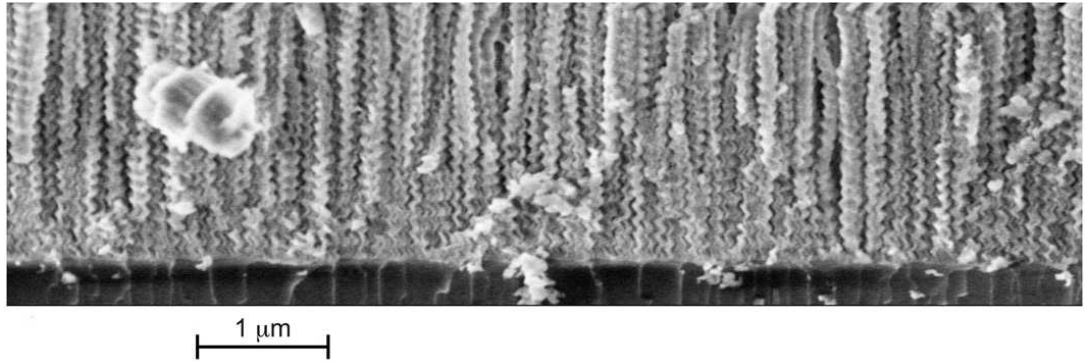}}
  \end{center}
  \caption{Scanning electron micrograph showing the helical nanowires of a chiral STF. Courtesy: M. W. Horn, The Pennsylvania State University.} \label{F: Example STF Microstructure}
\end{figure}

The curvilinear--nanowire morphology of STFs is exemplified by Fig.~\ref{F: Example STF Microstructure}. Note the individual nanowires that make up the STF in this case are helical; they are also nominally identical and parallel. The constituent nanowires of an STF have diameters of $\sim$10--300~nm, and lengths on the order of tens of nanometers to several micrometers. The nanowires are made of clusters 1--3~nm in linear dimension. Therefore, STFs may be classified as nanomaterials. For many applications dependent on optical or infrared radiation, the closely packed nanowires can be treated effectively as equivalent to an anisotropic and nonhomogeneous continuum.

% give brief history

Possibly the first STF was fabricated by Young and Kowal in 1959~\cite{N.O.Young-1959(P)}, who made fluorite films via oblique evaporation onto a rotating substrate. Although the structure of the resulting film was not unambiguously determined, the film did display optical activity. Most likely, the film was structurally chiral, as in Fig.~\ref{F: Example STF Microstructure}. This development was preceded by about 75 years of growing columnar thin films by thermal evaporation \cite{R.Messier-2008(P)}. Columnar thin films are assemblies of parallel straight nanowires, and are optically akin to biaxial crystals. The emergence of scanning electron microscopy in the late 1950s and early 1960s led to the confirmation of nanowire morphology. In 1966, a significant event occurred when Nieuwenhuizen and Haanstra showed that the nanowire inclination could be altered within a transitional distance of about 30~nm~\cite{J.M.Nieuwenhuizen-1966(P)}.

Messier and colleagues identified the conditions for the emergence of the nanowire morphology~\cite{R.Messier-1984(P)}. In 1989, Motohira and Taga showed that the direction of growing nanowires could be altered often within a few nanometers by appropriately switching the direction of the incoming vapor mid-deposition~\cite{T.Motohiro-1989(P)}. Lakhtakia and Messier realized that nanowires with almost arbitrary shapes could be fashioned by continuously changing the direction of the incoming vapor during deposition~\cite{A.Lakhtakia-1994p(P)}. Robbie and colleagues reported the first STF with unambiguously determined structure~\cite{K.Robbie-1995(P)}. Since that time, hundreds of research papers, a monograph~\cite{A.Lakhtakia-2005(B)}, and many book chapters have been written on STFs~\cite{V.C.Venugopal-2000(C), A.Lakhtakia-2003(C), J.B.GeddesIII-2006(C), J.A.Polo-2006(C), F.Wang-2007(C)}.

% object of this review

This article contains a brief description of and the physical phenomenons exhibited by STFs. In many cases, those phenomenons are due directly to the symmetries that the structures of the constituent nanowires break. 
In Section~\ref{S: Fabrication}, we review the fabrication of STFs. Then in Section~\ref{S: Constitutive Relations}, we present the mathematical description of the optical constitutive relations of STFs. In Section~\ref{S: Applications}, we review the optical, thermal, chemical, and biological applications to which STFs either have been put or have been suggested. The advantages and disadvantages of STF technologies compared to other technologies are presented in the same section. We end  in Section~\ref{S: Concluding Remarks} with some remarks on the extension of the STF concept for acoustic applications.

% and formulate the canonical wave--propagation problems for STFs in the frequency domain

% We then describe several selected examples in Section~\ref{S: Selected Examples} that illustrate both the wide range of applications of STFs, and how they are beginning to be incorporated in systems as opposed to one-off devices.

% Also, the examples are chosen to illustrate the connections between phase, length, and time in optics and the consequences of broken symmetries.

\section{Fabrication of STFs} \label{S: Fabrication}

% brief intro to fabrication

\begin{figure}[t]
  \begin{center}
    \scalebox{0.3}{\includegraphics{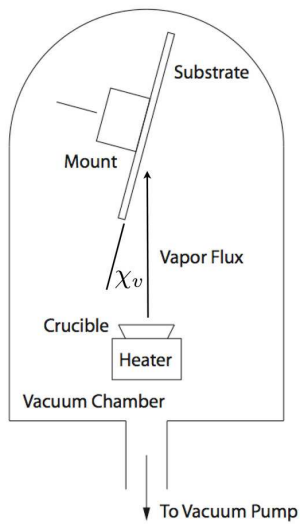}}
  \end{center}
  \caption{Schematic for fabrication of STFs by thermal evaporation. The source material is placed in a heated crucible, from which the vapor flux is allowed to fall obliquely on a substrate. A motorized mount allows the orientation of the substrate to be varied with time.} \label{F: Fabrication of STFs}
\end{figure}

STFs are chiefly fabricated by physical vapor deposition~\cite{R.Messier-2000(P)}. Several variants exist, of which thermal evaporation is the simplest. A source material is evaporated under high vacuum and the vapor allowed to flow at an oblique incidence angle $\chi_{v}$ onto a substrate, as schematically depicted in Fig.~\ref{F: Fabrication of STFs}. When the temperature of the source material is less than approximately one--third the melting temperature, there is little surface diffusion, and straight nanowires form. The morphology is enabled by the shadowing of nanowires by those in front of them.

\begin{figure}[htb]
  \begin{center}
    \scalebox{0.4}{\includegraphics{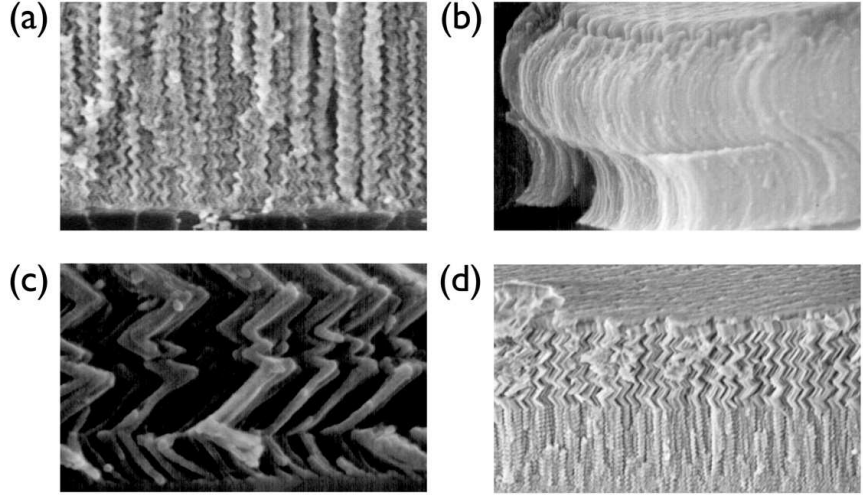}}
  \end{center}
  \caption{Examples of (a) thin--film helicoidal bianisotropic medium, (b) and (c) sculptured nematic thin films, and (d) hybrid STFs. Courtesy: M. W. Horn and R. Messier, The Pennsylvania State University.} \label{F: Types of STFs}
\end{figure}

The nanowire morphology can be changed during deposition by tilting and rotating the substrate. When the substrate is rotated about a direction perpendicular to the substrate and passing through it, the nanowires become helical. These structures form the basis of chiral STFs, also known as thin--film helicoidal bianisotropic mediums (TFHBMs) in optical parlance. Several scanning electron micrographs of STFs, including TFHBMs, are shown in 
Fig.~\ref{F: Types of STFs}.

In addition, the substrate can be rotated about an axis in the plane of the substrate. Changing the tilt along either or both of two such orthogonal axes can lead to zigzag, c--, and s--shaped nanowires. Such STFs are known as sculptured nematic thin films (SNTFs). Several scanning electron micrographs of SNTFs are shown in Fig.~\ref{F: Types of STFs}.

Hybrid STFs that consist of both TFHBM and SNTF sections grown one after another are also possible. For example, Suzuki and Taga fabricated hybrid STFs with zigzag, helicoidal, and straight nanowire sections~\cite{M.Suzuki-2001(P1)}. More recently, Park {\it et al.} grew a hybrid STF comprising three different chiral sections~\cite{Y.J.Park-2008(P)}.

Several variations on the basic scheme are possible, including serial bideposition, multideposition, prepatterning of substrates, and hybrid chemical/physical vapor deposition. In serial bideposition, a single source is used to fabricate films with large local birefringence~\cite{I.J.Hodgkinson-1999(P1), I.J.Hodgkinson-2000(P)}. For example, when growing chiral STFs, the source is manipulated to deposit alternately on either side of the growing nanowires. Multideposition consists of two or more sources being simultaneously evaporated to grow STFs~\cite{A.Lakhtakia-2008(P1)}. Hybrid chemical/physical vapor deposition has been used to fabricate polymeric STFs~\cite{S.Pursel-2005(P), M.C.Demirel-2007(P)}, because such materials cannot be heated to produce a collimated vapor.

  \begin{center}
\noindent{Table 1.} {Materials from which STFs have been fabricated. Note that the list includes oxides, fluorides, metals, semiconductors, and polymers. Materials marked with $^{*}$ indicate inverse structures formed by infilling of STF voids with the indicated material and subsequent removal of the initial STF.}\vskip 5pt
{\small    \begin{tabular}{||c|c||c|c||}
      \hline
      Material(s) & Reference(s) &  Material(s) & Reference(s) \\
      \hline
      iron oxide & \cite{H.Tan-2006(P)} &
      silicon oxide & \cite{R.Messier-2000(P), K.Robbie-1996(P), M.W.Horn-2004(P1), M.W.Horn-2004(P2)} \\
      tantalum oxide & \cite{I.Hodgkinson-1998(P)} &
      tin oxide & \cite{M.W.Horn-2004(P1)} \\
      titanium oxide & \cite{I.J.Hodgkinson-2000(P), I.Hodgkinson-1998(P), A.C.vanPopta-2004(P)} &
      zirconium oxide & \cite{I.Hodgkinson-1998(P), K.D.Harris-2001(P)} \\
       calcium fluoride & \cite{K.Robbie-1995(P), K.Robbie-1996(P)} &
      magnesium fluoride & \cite{K.Robbie-1995(P),  R.Messier-2000(P), K.Robbie-1996(P)} \\
        indium nitride & \cite{Y.Inoue-2007(P)} &
       aluminum & \cite{M.W.Horn-2004(P1)} \\
       chromium & \cite{K.Robbie-1996(P), M.W.Horn-2004(P1)} &
      copper & \cite{K.Robbie-1996(P)} \\
      gold & \cite{A.L.Elias-2004(P)}$^{*}$ &
       manganese & \cite{K.Robbie-1996(P), J.N.Broughton-2002(P)} \\
      molybdenum & \cite{M.W.Horn-2004(P1)} &
      nickel & \cite{A.L.Elias-2004(P)}$^{*}$ \\
       carbon & \cite{G.K.Kiema-2003(P)} &
       silicon & \cite{E.Schubert-2006(P), D.X.Ye-2005(P), D.X.Ye-2007(P)} \\
      GeSbSe chalcogenide & \cite{R.J.MartinPalma-2007(P1), R.J.MartinPalma-2007(P2), R.J.MartinPalma-2008(P)} &
      Alq$_{3}$ & \cite{P.C.P.Hrudey-2006(P)} \\
      parylene & \cite{S.Pursel-2005(P), M.C.Demirel-2007(P)} &
      polystyrene & \cite{A.L.Elias-2004(P)}$^{*}$ \\
%      teflon & \cite{} \\
      \hline
    \end{tabular}
    }
  \end{center}

Inverse structures can be formed by infilling of STF voids via electrodeposition of metal or infiltration of molten polyer, and subsequent removal of the initial STF via etching~\cite{A.L.Elias-2004(P)}.

STFs can be fabricated from just about any material that can be evaporated. A wide variety of materials, including dielectrics, metals, and polymers have been used to fabricate STFs. Table~1 lists several materials from which STFs have been fabricated. GeSbSe chalcogenide glasses and luminescent Alq$_{3}$ have been used most recently~\cite{R.J.MartinPalma-2007(P1), R.J.MartinPalma-2007(P2), R.J.MartinPalma-2008(P), P.C.P.Hrudey-2006(P)}.

The structures into which STFs can be formed has also expanded to include three--dimensional STF architectures. STFs can be deposited on substrates that have been prepatterned via lithography or other processes. The STF nanowires grow on the raised areas (and, depending on the angle of incidence of the vapor, the side walls) to form three--dimensional structures~\cite{M.W.Horn-2004(P1), M.W.Horn-2004(P2)}. Alternatively, the pre--deposition of arrays of small islands, or seeds, of source material can result, after STF deposition, in patterns of single nanowires or clusters of nanowires~\cite{M.Malac-1999(P), B.Dick-2003(P), M.O.Jensen-2005(P)}. 

% mention GLAD

%For more detailed information on the fabrication of STFs, we refer readers to several review articles~\cite{R.Messier-2000(P), I.Hodgkinson-2001(P)} and a book on the subject~\cite{A.Lakhtakia-2005(B)}, as well as the primary literature.

\section{Optical Constitutive Relations of STFs} \label{S: Constitutive Relations}

Consider an STF that occupies the region $0 \leq z \leq L$. We use the cartesian coordinate system $\L( x, y, z \R)$ with the triad of unit vectors $\L\{ \UV{x}, \UV{y}, \UV{z} \R\}$. The frequency-domain constitutive relations of such a dielectric, spatially local STF within this region are as follows:
\begin{eqnarray}
\DF \XYZW & = & \Eo \, \EE \iv{z, \FQ} \IP \EF \XYZW \, , \label{E: Electric induction} \\
\BF \XYZW & = & \Mo \, \HF \XYZW \, , \label{E: Magnetic induction}
\end{eqnarray}
where $\EF$ and $\BF$ are the macroscopic primitive electric and magnetic field phasors; $\DF$ and $\HF$ are the corresponding induction field phasors; $\Eo$ and $\Mo$ are the permittivity and permeability of vacuum, respectively; $\FQ$ is the angular frequency; and $\EE$ is the relative permittivity dyadic. The latter can be factored into pieces which depend either only on the structure of the film or the dispersive properties; thus,
\begin{equation}
\EE \iv{z, \FQ} = \RD{} \iv{z} \IP \ER \iv{\FQ} \IP \RD{} \iv{z} \, ,
\end{equation}
where $\ER$ is the reference permittivity dyadic, so named to correspond to a reference plane $z = z_{o}$ where the rotation dyadic $\RD{}$ equals the identity dyadic (i.e., $\RD{} \iv{z_{o}} = \ID$).

There are two canonical forms for the rotation dyadic. When
\begin{equation}
\RD{} \iv{z} = \RD{y} \iv{z} = \UV{y} \UV{y} + \L( \UV{x} \UV{x} + \UV{z} \UV{z} \R) \cos \tau \iv{z} + \L( \UV{z} \UV{x} - \UV{x} \UV{z} \R) \sin \tau \iv{z} \, ,
\end{equation}
the STF exhibits an essentially two--dimensional morphology (in the $xz$ plane), and hence is called an SNTF. The angle $\tau \iv{z}$ determines whether the nanowires are chevrons, c--shaped, s--shaped, or of other more complex shapes.

When
\begin{equation}
\RD{} \iv{z} = \RD{z} \iv{z} = \UV{z} \UV{z} + \L( \UV{x} \UV{x} + \UV{y} \UV{y} \R) \cos \zeta \iv{z} + \L( \UV{y} \UV{x} - \UV{x} \UV{y} \R) \sin \zeta \iv{z} \, ,
\end{equation}
the STF exhibits a three--dimensional morphology, and hence is called a TFHBM. Such films are structurally chiral. Often such STFs are formed with $\zeta \iv{z} = \pi z / \Omega_{c}$ and are therefore periodic, where $\Omega_{c}$ is the structural half--period. These periodically nonhomogeneous STFs are called chiral STFs.

In more complicated STFs, $\RD{} \iv{z}$ can be some combination of $\RD{y} \iv{z}$ and $\RD{z} \iv{z}$. In hybrid STFs, both $\ER \iv{\FQ}$ and $\RD{} \iv{z}$ can have sectionwise variations along the $z$ axis.

A link between the microscopic structure of an STF and its macroscopic constitutive properties is found by local homogenization in a nominal model of STF morphology~\cite{J.A.Sherwin-2001(P)}. This nominal model considers each STF nanowire as being composed of a string of homogenous ellipsoids; the voids are also treated as conglomerates of ellipsoids. Local homogenization of this model, for example with the Bruggeman formalism, yields $\ER \iv{\FQ}$. An extended model shows how STFs infiltrated with a chiral fluid could exhibit bianisotropic optical properties~\cite{J.A.Sherwin-2002(P), J.A.Sherwin-2003(P)}. 

\section{Applications of STFs} \label{S: Applications}

% optical
% thermal
% dielectric / electronic
% chemical
% biological

A variety of the optical, electronic, chemical, and biological applications of STFs have been proposed and demonstrated.

\subsection{Optical} \label{SS: Optical Applications}

The area of application for which STFs have been developed the most thoroughly is optics. The nanometer--scale control over the structure of STFs has meant that the optical properties of the films could be controlled on subwavelength scales. The recent development of three--dimensional STF architectures means that simultaneous control at both wavelength and subwavelength scales is possible---as in three--dimensional photonic crystals.

Even within the optics category, the most well--developed area is filter technology, particularly that relying on the circular Bragg phenomenon exhibited by chiral STFs. The chirality and periodicity of the latter can give rise to a stop band along the thickness direction. That is, left/right circularly polarized light over a bandwidth called the Bragg regime is mostly reflected by a sufficiently thick chiral STF that is structurally left/right handed, but right/left circularly polarized light is either transmitted or absorbed. In a more abstract sense, the chiral STF breaks both rotational and translational symmetries, and this gives rise to the circular Bragg phenomenon.

Due to their controllable void fraction, STFs have been suggested as low--permittivity dielectrics to reduce crosstalk in integrated circuits~\cite{V.C.Venugopal-2000(P)}. Also, efforts have begun to integrate active light emission and gain elements into STFs for optoelectronic applications. For example, a layer of optically pumped quantum dots was embedded between two chiral STF layers to create a device which emitted circularly polarized light~\cite{F.Zhang-2007(P)}. Chiral STFs were also used to create a circularly polarized external--cavity diode laser~\cite{F.Zhang-2008(P)}. In addition, the possibility of tunable polymeric chiral STF lasers, whose microstructure can be compressed and released by a piezoelectric element to control the output wavelength, has been analyzed~\cite{F.Wang-2002(P)}. A similar scheme has been suggested for tunable optical filters~\cite{F.Wang-2003(P)}.

Another route to tunability is to infiltrate the pores of, for example, chiral STFs with liquid crystals. The aciculate molecules of the liquid crystal tend to align with the STF nanowires; this effect has been shown to increase the optical rotary power of the resulting composite films~\cite{K.Robbie-1999(P)}. Moreover, one can apply an electric field to align the liquid crystal molecules along the field, with a concomitant change in optical properties for the composite~\cite{J.C.Sit-2000(P)}.
 
A third route to tunability is to fabricate STFs from electro--optic materials~\cite{A.Lakhtakia-2008(P1), A.Lakhtakia-2008(P2), A.Lakhtakia-2008(P3)}. Most notably, the chiral STFs exhibiting the Pockels effect can exhibit a Bragg regime in the presence of a low--frequency electric field~\cite{J.A.Reyes-2006(P)}.

A major emerging area in optics is to exploit STFs for plasmonics--based sensing of chemical and biological materials. For example,
ultrathin metal STFs have shown a sensitivity that is significantly larger than for bulk films: $113^\circ$/RIU
 for metal STFs in contrast to $79^\circ$/RIU for bulk films \cite{A.Shalabaney-2009(P)}. Most recently, the propagation
of multiple surface--plasmon--polariton waves, of the same frequency but different phase speeds, guided
by a planar metal/STF interface has been theoretically predicted
\cite{M.A.Motyka-2008(P),M.A.Motyka-2009(P),J.A.Polo-2009(P1),J.A.Polo-2009(P2)}  and experimentally verified
\cite{A.Lakhtakia-2009(P)}, with exciting prospects for multianalyte sensing and error--free sensing.

Regardless of the excellence of optical performance characteristics, no STF can be used reliably if it
cannot withstand mechanical stimuli within reasonable ranges. Various mechanical properties of STFs,
such as hardness, compliances, yield stress, and ultimate strength, need to be modeled and measured in relation to
STF morphology and composition. Furthermore, there is no experimental evidence showing how the
optical  properties change with the application of stress. This is practically virgin
territory for research. It is known from indentation experiments that chiral STFs are permanently
deformed when applied pressures exceed 30~$\mu$Pa \cite{M.W.Seto-1999(P)}. This problem
could be alleviated by impregnating an STF with a polymer \cite{A.L.Elias-2004(P),V.C.Venugopal-2000(P)} which shall have to be accounted
for during optical design; however, that step will require experimentation. Moreover,
polymer infiltration would allow STFs to be used as optical pressure sensors \cite{A.Lakhtakia-2005(B)} as well as further the
development of mechanically tunable light emitters and optical filters \cite{F.Wang-2002(P),F.Wang-2003(P)}.

\subsection{Thermal} \label{SS: Thermal Applications}

Harris {\it et al.}\ deposited alternating layers of solid and porous SNTFs of yttria--stabilized zirconia~\cite{K.D.Harris-2001(P)}. They found that the composite structure had a reduced thermal diffussivity compared to bulk films, with the added benefit of providing stress relief from expansion. The particular form of the SNTF structure they used was required to prevent cracking in the films.

\subsection{Chemical} \label{SS: Chemical Applications}

A variety of photochemical, electrochemical, and fluid sensor applications of STFs have also begun to be investigated. 

Titanium dioxide STFs have been integrated into Gr\"{a}tzel cells for solar--power applications~\cite{S.M.Pursel-2007(P)}. The photocatalytic properties of titanium--dioxide STFs have been found to depend on their morphology~\cite{M.Suzuki-2001(P1)}.

The porous structure of indium--nitride SNTFs and TFHBMs was shown to improve their electrochromic properties~\cite{Y.Inoue-2007(P)}. In addition, Brett and colleagues measured the electrohemical properties of carbon TFHBMs and chevronic manganese SNTFs~\cite{J.N.Broughton-2002(P), G.K.Kiema-2003(P)}.

The porosity of STFs means that chemicals can infiltrate their pores and change their optical and electrical properties. For example, by depositing titanium--dioxide STFs on coplanar interdigitated electrodes, sensors whose capacitance increases with increasing relative humidity can be made~\cite{J.J.Steele-2007(P)}. Both the optical rotation and location of spectral hole within the circular Bragg phenomenon of a chiral STF also respond to the relative humidity of the environment~\cite{I.J.Hodgkinson-1999(P2), J.J.Steele-2006(P)}. Shifts in spectrums have also been observed on complete infiltration of the voids in chiral STFs by acetone, methyl alcohol, isopropyl alcohol, and hexamethyldisilizane~\cite{S.M.Pursel-2007(P)}. This experimental work was anticipated by earlier theoretical studies of both SNTFs and TFHBMs as fluid sensors~\cite{E.Ertekin-1999(P), A.Lakhtakia-2001(P1), A.Lakhtakia-2001(P2)}. The susceptibility of the optical properties of STFs to ambient moisture underscores the need for the proper packaging of films not intended for use as sensors.

Although most STFs tend to absorb ambient moisture, they can under certain conditions exhibit superhydrophobic behavior, especially when functionalized with siloxane~\cite{S.Tsoi-2004(P), N.Verplanck-2007(P)}. Fluorination should also promote superhydrophobicity~\cite{Y.Zhou-2006(P)}.

\subsection{Biological} \label{SS: Biological Applications}

These chemical applications are quickly being joined by biological ones. Magnesium--alloy STF coatings on stents are being fabricated to be bioabsorbable~\cite{S.M.Pursel-2007(P)}. Parylene STFs have been fabricated as scaffolds on which, for example, both kidney and fibroblast cells can grow~\cite{A.Lakhtakia-2008(P1), M.C.Demirel-2006(P)}. STFs are particularly suited to serve as scaffolds because their structure effectively mimics that of a tissue's extracellular matrix~\cite{A.Lakhtakia-2008(P1)}. The ability to make transparent STFs means that cell cultures grown within them can be imaged, for example with confocal microscopy~\cite{A.Lakhtakia-2008(P1)}. The degradation of iron(III) oxide STFs by metal--reducing bacteria has been shown to affect their optical properties, a result that indicates the possibility of using STF-based fiber optic sensors to measure bacterial activity below ground~\cite{H.Tan-2006(P)}.

\subsection{Advantages and Disadvantages} \label{SS: Advantages and Disadvantages}

STF technology presents several advantages as compared with competing technologies.
The fabrication process of STFs is very simple and cheap, especially as compared to conventional semiconductor patterning technologies.
The fabrication process affords a high degree of control over STF morphology, and consequently properties. 
STFs can be made of many different kinds of materials.
However, there are disadvantages too.
STFs can be fragile, and so must be suitably packaged to prevent damage.
Suitable packaging is also needed to prevent absorption of moisture and concomitant change in properties, unless the films are being used as sensors.

Thus STFs are probably most appropriate for applications where a high degree of control over film properties and cost are important, but mechanical stability is of lesser concern.

\begin{figure}[t]
  \begin{center}
    \scalebox{0.8}{\includegraphics{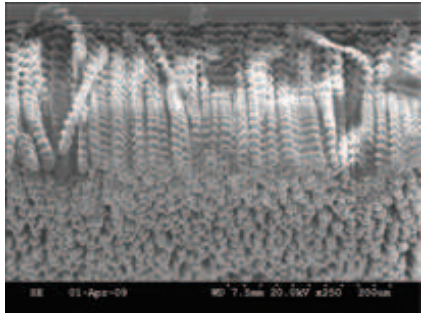}}
  \end{center}
  \caption{Scanning electron micrograph of a chiral STF of parylene fabricated for acoustic purposes.} \label{F: Example STF Acoustics}
\end{figure}

\section{Concluding Remarks} \label{S: Concluding Remarks}

Extension of the STF concept for acoustic purposes requires a different length scale than that for optical purposes:
nanometers have to be replaced by micrometers. Very recently, polymer STFs have been made on the
microscale by a hybrid vapor deposition process; one such STF is shown in Fig.~\ref{F: Example STF Acoustics}. Such films can
be used as acoustic filters, both for polarization and bandwidth 
\cite{A.Lakhtakia-1994(P),A.Lakhtakia-1997(P),A.Lakhtakia-1999(P2),A.Lakhtakia-2000(P),R.J.Carey-2006(P)}. These films could also be used
as MEMS elements, particularly for sensing. The acoustic properties of these STFs require
experimental research coupled with theoretical modeling \cite{A.Lakhtakia-2000(P2),A.Lakhtakia-2001(P3),A.Lakhtakia-2002(P)}.

To conclude, STFs show promise for several optical, electronic, chemical, biological, and acoustic applications. Of these, the optical applications are the most developed. The controllable morphology and wide variety of constituent materials makes these thin-film metamaterials attractive for these and other applications.

%------------------------------%
% End matter:
%------------------------------%
\vskip 8pt
\noindent {\small {\bf Acknowledgments:} J.\ B.\ Geddes III gratefully acknowledges support of a Beckman Postdoctoral Fellowship. A.\ Lakhtakia thanks the Binder Endowment at Penn State for financial support of his research.}

\bibliography{Lakhtakia_09_6arx}

\end{document}